\documentclass[12pt,notitlepage,a4paper]{article}
\pdfoutput=1

\usepackage{epsfig}
\usepackage{color,graphicx}
\usepackage{cite}
\usepackage{amssymb,amsmath}

\newcommand{\be}{\begin{equation}}
\newcommand{\ee}{\end{equation}}
\newcommand{\bea}{\begin{eqnarray}}
\newcommand{\eea}{\end{eqnarray}}

\begin{document}

\begin{flushright}
CCTP-2012-23 
\end{flushright}

\begin{center}  

\vskip 2cm 

\centerline{\Large {\bf Fluctuations and instabilities of a holographic metal}}
\vskip 1cm

\renewcommand{\thefootnote}{\fnsymbol{footnote}}

\centerline{Niko Jokela,${}^1$\footnote{niko.jokela@usc.es}
Matti J\"arvinen,${}^2$\footnote{mjarvine@physics.uoc.gr} 
and Matthew Lippert${}^{2,3}$\footnote{M.S.Lippert@uva.nl}}

\vskip .5cm
${}^1${\small \sl Departamento de F\'isica de Part\'iculas} \\
{\small \sl Universidade de Santiago de Compostela}\\
{\small \sl and}\\
{\small \sl Instituto Galego de F\'isica de Altas Enerx\'ias (IGFAE)}\\
{\small \sl E-15782 Santiago de Compostela, Spain} 

\vskip 0.2cm
${}^2${\small \sl Crete Center for Theoretical Physics} \\
{\small \sl Department of Physics} \\
{\small \sl University of Crete} \\
{\small \sl 71003 Heraklion, Greece} 

\vskip 0.2cm
${}^3${\small \sl Institute for Theoretical Physics} \\
{\small \sl University of Amsterdam} \\
{\small \sl 1090GL Amsterdam, Netherlands} 

\end{center}

\vskip 0.3 cm

\setcounter{footnote}{0}
\renewcommand{\thefootnote}{\arabic{footnote}}

\begin{abstract}
\noindent  
We analyze the quasinormal modes of the D2-D8' model of 2+1-dimensional, strongly-coupled, charged fermions in a background magnetic field and at non-zero density.  The model is known to include a quantum Hall phase with integer filling fraction.  As expected, we find a hydrodynamical diffusion mode at small momentum and the nonzero-temperature holographic zero sound, which becomes massive above a critical magnetic field.  We confirm  the previously-known thermodynamic instability.  In addition, we discover an instability at low temperature, large mass, and in a charge density and magnetic field range near the quantum Hall phase to an inhomogeneous striped phase. 

\end{abstract}

\newpage

\tableofcontents

\section{Introduction}
\label{sec:intro}

Understanding the behavior of strongly interacting electrons in condensed matter systems is a long-standing and difficult problem.  Among gapless systems with unbroken translation and charge conservation symmetry, Landau-Fermi liquids are the one well-understood class; much less is known about compressible systems that are not continuously connected to free fermions.  Non-Fermi liquids, such as strange metals, which have Fermi surfaces but exhibit very different low-energy behavior, have been a hot topic recently, but they remain essentially mysterious.  There is also the controversial possibility of compressible states with unbroken symmetry that are neither Fermi liquids nor non-Fermi liquids because they lack Fermi surfaces altogether.

Studying the possible compressible states of fermionic matter would appear to be a problem amenable to gauge/gravity duality.  Holography provides tractable descriptions of strongly-coupled systems and elegantly avoids the fermion sign problem typically plaguing calculations at nonzero density. 

We will take a top-down approach, with the goal of constructing a broad class of related string theory models and investigating both their common, universal features and the range of possible phenomena.  Perhaps the most tractable and well-studied such group is the family of D$p$-D$q$ brane intersection models, which includes, for example, the supersymmetric D3-D7 \cite{Karch:2002sh}, the related non-supersymmetric D3-D7' \cite{Bergman:2010gm}, the Sakai-Sugimoto \cite{Sakai:2004cn}, and the D3-D5 models \cite{D3D5}, as well as many other variants in 1+1, 2+1, and 3+1 dimensions.\footnote{For a partial list, see \cite{otherDpDq}.}  These brane systems describe an unusual compressible phase, the holographic quantum liquid; despite compelling arguments that a Fermi surface ought to exist \cite{sachdev}, various attempts\footnote{Bottom-up investigations of holographic matter have uncovered many phenomena which are intimately related to the existence of the Fermi surface.  Two recent examples are Friedel oscillations \cite{Puletti:2011pr} and Landau levels \cite{Blake:2012tp}.} have failed to find convincing evidence of one \cite{Karch:2008fa, kp, Ammon:2011hz, Goykhman:2012vy, Brattan:2012nb}.   It is currently an open problem whether the Fermi surface is really absent or whether it is fractionalized or otherwise hidden from the view of gauge invariant observables.  

One interesting new member of this family of holographic quantum liquids is the D2-D8' system \cite{Jokela:2011eb},  a holographic model for 2+1-dimensional, strongly-coupled fermions.  This system, a close cousin of the D3-D7' model\footnote{Also, see \cite{D3-D7'precursors,Myers:2008me} for earlier developments leading up to these types of models.} \cite{Bergman:2010gm}, consists of a stabilized probe D8-brane in the background of a stack of D2-branes such that the number of dimensions with mixed Neumann-Dirichlet boundary conditions \#$ND = 6$.  Unlike the D3-D7' model, the D2-D8' system exhibits hyperscaling violation, with an exponent $\theta = -\frac{1}{3}$ \cite{Dong:2012se}.  This is simply a consequence of the fact that the coupling constant is running trivially. Indeed, one can trace back the conformal behavior in effectively higher dimensions by switching to a dual frame \cite{Kanitscheider:2008kd}.  At nonzero charge density and magnetic field, the spectrum contains a gapped quantum Hall state with filling fraction one, whose excitations were analyzed in \cite{Jokela:2011sw} and where, in particular, the existence of the magneto-roton as the lowest excitation mode was established.  In this paper, we focus instead on excitations of the gapless, metallic states.  

We find a pattern of low-lying quasi-normal modes qualitatively similar to that found in other D$p$-D$q$ systems \cite{Karch:2008fa, kp, Kim:2008bv, Kaminski:2009dh, Bergman:2011rf, Davison:2011ek, Jokela:2012vn, Goykhman:2012vy, Brattan:2012nb,Gorsky:2012gi}.  At long wavelength the longest lived mode is, as expected, a diffusion mode associated with charge conservation.  At shorter wavelength, crossing over from the hydrodynamic to collisionless regimes, the diffusion mode merges with another purely dissipative mode, and the holographic zero sound mode appears.  Turning on a magnetic field rearranges the spectrum of quasinormal modes and, when sufficiently large, gives a mass to the zero sound.

Unlike the stable quantum Hall state, the metallic state is unstable in large regions of parameter space.  We find two types of instability: a thermodynamic instability, which occurs next to the quantum Hall state, and spatially modulated instability at low temperature and for magnetic fields roughly of the same size as the charge density.  The first of these was observed in \cite{Jokela:2011eb} as a negative charge susceptibility.  Here, we see this same instability manifest in the linear fluctuation analysis as a diffusion mode with negative diffusion constant.  

The modulated instability is induced by a nontrivial term in the D8-brane Chern-Simons action and is in the family of instabilities seen first in \cite{Nakamura:2009tf}.  Instabilities at nonzero momentum due to Chern-Simons terms turn out to be quite generic and have been found recently in a variety of holographic systems \cite{inhomo, Bergman:2011rf, Jokela:2012vn}.  We have not yet found the inhomogeneous ground state which follows from this instability, because the worldvolume fields on the D8-brane are fully coupled. We anyhow expect that it will be a striped phase with all quantities, such as spin, charge density, and transverse currents, spatially modulated.

Although our results are in many ways qualitatively similar to those found in the D3-D7 and D3-D7' models (for a recent review, see \cite{Bergman:2012na}), there are some differences worth highlighting.  The D3-D7 model is supersymmetric and free of instabilities \cite{Ammon:2011hz}.  The D3-D7' system, on the other hand, suffers an analogous modulated instability to the one found here.  In D3-D7' model, increasing the charge density exacerbated the modulated instability, while adding a magnetic field or increasing the mass improved stabity, in line with expectations from the original supergravity model \cite{Nakamura:2009tf}.  Surprisingly, we find that the D2-D8' system is unstable near the quantum Hall state, that is, when the charge and magnetic field are roughly equal.  When either the charge or magnetic field is much larger than the other, the homogeneous state becomes stable.  Also, in contrast with the D3-D7' result, increasing the mass enhances the instability.  Furthermore, the thermodynamic instability observed earlier in \cite{Jokela:2011eb} near the quantum Hall state is present only in D2-D8' model and is found to be continuously connected with the modulated instability.

Although the D3-D7' model also features a stable quantum Hall state, it was not included in the parameter space explored in \cite{Bergman:2011rf, Jokela:2012vn}.  Here we see for the first time how stability is recovered as the gapped, quantum Hall state is approached:  rather than the range of unstable momenta shrinking, whole range runs off to infinity as the mass gap opens.  

We begin in Sec.~\ref{sec:review} by reviewing the D2-D8' model, its construction and background solutions.  Then, in Sec.~\ref{sec:fluctuations} we set up the fluctuation analysis of the metallic phase.  Sec.~\ref{sec:results} contains our results on the quasinormal mode spectrum and instabilities, and we conclude in Sec.~\ref{sec:discussion}.

\section{Review of the D2-D8' model}
\label{sec:review}
The D2-D8' system, introduced in \cite{Jokela:2011eb}, consists of a probe D8-brane in the near-horizon background of $N$ near-extremal D2-branes, such that in the flat-space limit $\# ND=6$.  We consider the limit of a large number of D2-branes $N \gg 1$ and large 't Hooft coupling $g_s N \gg 1$. Supersymmetry is completely broken, and internal flux is needed for stability.  At weak coupling, the low-energy spectrum of bifundamental strings in a $\# ND=6$ system contains only charged fermions and no charged bosons; as a result, the holographic field theory dual is 2+1-dimensional SYM coupled to $N$ species of charged fermions.  Similar constructions have been used in 2+1-dimensional D3-D7 intersections \cite{Myers:2008me, Bergman:2010gm}.

\subsection{Set-up}
The ten-dimensional near-horizon metric for a thermal D2-brane is \cite{Itzhaki:1998dd}
\be
L^{-2} ds^2_{10} = u^\frac{5}{2}\left(-f(u)dt^2+dx^2+dy^2\right)+u^{-\frac{5}{2}}\left(\frac{du^2}{f(u)}+ u^2 d\Omega_{S^6}^2\right) \ ,
\ee
where $L^5 = 6\pi^2 g_s N l_s^5$ and the thermal factor $f(u)=1-\left(\frac{u_T}{u}\right)^5$.  Here we use dimensionless (lowercase) coordinates, which are related to the physical dimensionful (uppercase) coordinates as: $x^\mu = X^\mu/L$, $u = U/L$.   
In general, lowercase Latin letters will denote dimensionless quantities, and uppercase letters will denote the corresponding physical quantities.  
However, for the D8-brane worldvolume gauge field $A_\mu$, we will define a dimensionless gauge field $a_\mu = \frac{2\pi\alpha'}{L} A_\mu$.

There is a black hole horizon at $u = u_T$, and the corresponding Hawking temperature is $T= \frac{5}{4\pi L} u_T^{3/2}$.  The metric on the $S^6$ is given by
\be
d\Omega_{S^6}^2 = d\psi^2 + \sin^2\psi \left(d\theta_1^2 + \sin^2\theta_1 d\phi_1^2\right) + \cos^2\psi \left(d\xi^2 + \sin^2\xi d\theta_2^2+ \sin^2\xi \sin^2\theta_2 d\phi_2^2\right) \ ,
\ee
where $\psi$ ranges from 0 to $\pi/2$, $\xi$, $\theta_1$, and $\theta_2$ range from 0 to $\pi$, and $\phi_1$ and $\phi_2$ range from 0 to $2\pi$. The background dilaton and RR 5-form potential are
\bea
e^\phi &=& g_s\left(\frac{L}{U}\right)^\frac{5}{4} \\
C^{(5)} &=& c(\psi) L^5  d\Omega_{S^2} \wedge d\Omega_{S^3} \ ,
\eea
where $c(\psi) = \frac{5}{8}\left(\sin\psi - \frac{1}{6}\sin(3\psi)- \frac{1}{10}\sin(5\psi)\right)$.

In this background, we consider a D8-brane probe filling the spacetime directions $t$, $x$, $y$, and the radial direction $u$.  In addition, the D8-brane wraps an $S^2 \times S^3$ fibered over an interval in the internal $S^6$.  We will take the ansatz that the embedding in the $\psi$-direction is a function only of $u$.  The induced metric on the D8-brane is
\bea
L^{-2}ds^2_{D8} &=& u^\frac{5}{2}\left(-f dt^2+dx^2+dy^2\right)+u^{-\frac{5}{2}}\left(\frac{1}{f} +u^2{\psi'}^2\right)du^2 \nonumber\\
&&+ u^{-\frac{1}{2}}\sin^2\psi \ d\Omega_2^2 + u^{-\frac{1}{2}}\cos^2\psi \ d\Omega_{3}^2 \ ,
\eea
where $' \equiv \partial_u$.

We will also turn on the following three components of the D8-brane worldvolume gauge field:
\bea
2\pi\alpha' F_{xy} &=& h \\
2\pi\alpha' F_{u0} &=& a_0'(u) \\
2\pi\alpha'F_{\theta_1\phi_1} &=& L^2 b \sin \phi_1 \ .
\eea
The constant spatial field strength corresponds to a background magnetic field $h$, the radial electric field is holographically dual to a charge density, and the magnetic field wrapping the internal $S^2$ is needed to stabilize the embedding.  Furthermore, we choose to work in the radial gauge, where $a_u = 0$.

The DBI action for the D8-brane is
\bea
 S_{DBI} &=&  - \mu_8 \int d^9x \, e^{-\phi} \sqrt{-\det(g_{\mu\nu} + 2\pi\alpha' F_{\mu\nu})}  \\
&=&  - \mathcal N \int du  \, u^\frac{5}{2}  \sqrt{A G H} \quad ,
\label{DBIaction}
\eea
where $\mathcal N = 8 \pi^3 T_8 v_3 L^9$ and $v_3$ is the dimensionless volume of spacetime.  The functions $A(u)$, $G(u)$, and $H(u)$ are defined as
\bea
A(u) &=& 1+ fu^2{\psi'}^2 - {a_0'}^2 \\
G(u) &=&  \cos^6\psi \left(b^2u + \sin^4\psi\right) \\
H(u) &=& 1+\frac{h^2}{u^5} \quad .
\eea

The Chern-Simons action has a nonzero term due to the charge density and magnetic field:
\be\label{CSaction}
S_{CS} = -\frac{T_8}{2}(2\pi\alpha')^2\int C^{(5)}\wedge F\wedge F = \mathcal N \int du \, c(\psi) ha_0' \ .
\ee

\subsection{Background solutions}
\label{sec:background}
The D8-brane action (\ref{DBIaction}) and (\ref{CSaction}) depends on $a_0$ only through its derivative.  We can then integrate the equation of motion for $a_0$ to obtain
\be
\frac{g}{f} H a_0' = d-hc(\psi) \equiv \tilde d(u) \ ,
\ee
where we have defined
\be
\label{ga0}
g(u) = f u^{5/2} \sqrt{\frac{G}{A H}} 
\ee
and where $d$ is the constant of integration and $\tilde d(u)$ is the radial displacement field.  In terms of the boundary theory, $d$ is the total charge density, while $\tilde d(u)$ is the charge due to sources located in the bulk at radial positions below $u$.

We can solve for $a_0'$ in terms of $\tilde d$, obtaining
\be
a_0' = \tilde d \sqrt{\frac{1+fu^2\psi'^2}{\tilde d^2 + u^5 G H}} \quad .
\ee
We then rewrite (\ref{ga0}) as
\be
\label{g}
g = \frac{f}{H} \sqrt{\frac{\tilde d^2 + u^5 G H}{1+f u^2\psi'^2}} \quad .
\ee

The $\psi$ equation of motion is given by
\bea
\label{psieom}
\partial_u\left(u^2 g H \psi'\right) &=& \frac{f u^5}{2g} \partial_\psi G - h a_0'  \partial_\psi c(\psi) .
\eea
We can solve (\ref{psieom}) at large $u$ and find that $\psi \to 0$ in the UV.  We identify the coefficient of the leading term in the asymptotic expansion of $\psi$ as the fermion mass and the coefficient of the subleading term as the chiral condensate: 
\be
\psi = \frac{m}{u} - \frac{c}{u^3} + \mathcal O(u^{-4}) \ .
\ee

There are two classes of D8-brane embeddings with different IR behavior: black hole (BH) embeddings, which enter the horizon at $u_T$, and Minkowski (MN) embeddings, which smoothly cap off at some $u_0 > u_T$ as the wrapped $S^3$ shrinks to zero size.  The DC conductivity can be computed straigthforwardly using the Karch-O'Bannon method \cite{KOB} in the black hole phase and a generalization thereof \cite{Bergman:2010gm} for the Minkowski 
phase.\footnote{Alternately, one can use either the Kubo formula or the open string metric method \cite{Kim:2011zd}.}  As argued in \cite{Jokela:2011eb}, the MN embeddings holographically reproduce the properties of a gapped, quantum Hall fluid.   In our previous paper \cite{Jokela:2011sw}, we analyzed the fluctuations of the MN solutions.  

Here we will focus on the quasi-normal mode excitations of the BH embeddings.  The BH solutions correspond to an ungapped, metallic phase known as a holographic quantum liquid, with several unusual properties.  At low temperature and no magnetic field, the longitudinal resistivity scales as $\rho_{xx} \sim T^{5/3}$ \cite{Karch:2009eb, Jokela:2011eb}, rather than like $T^2$ as for a Fermi liquid.  Such a fractional scaling exponent is due to hyperscaling violation; in a 2+1-dimensional system, thermodynamic quantities scale with an effective spatial dimension $2-\theta$, where $\theta$ is the hyperscaling violation exponent.  For a D$p$-brane background, $\theta = \frac{(p-3)^2}{p-5}$, which for D2-branes gives $\theta = -\frac{1}{3}$ \cite{Dong:2012se}.  No concrete evidence of a Fermi surface has been seen for this, or any other D$p$-D$q$ brane system \cite{Karch:2008fa, kp, Ammon:2011hz, Goykhman:2012vy, Brattan:2012nb}, though it has been argued that some sort of hidden fractionalized Fermi surface exists \cite{sachdev}.

The MN embeddings exist only when $d=\frac{3}{2} h$, meaning that the quantum Hall state is at filling fraction $\nu = 1$.  In \cite{Jokela:2011eb}, we found two types of transitions from the quantum Hall state to the metallic state.  As the temperature is increased, there is a first-order phase transition from the MN embedding to a BH embedding.  It persists for $d/h \approx \frac{3}{2}$ as a 
first-order transition between two different BH embeddings.  In this paper, we work at sufficiently low temperature such that we will completely avoid this transition.

The second kind of transition occurs when the charge or magnetic field is changed so that $d \ne \frac{3}{2} h$.  This causes the D8-brane to transition from a MN embedding to a BH embedding.  We found in \cite{Jokela:2011eb} that this process is not symmetric: increasing $d$ leads to a continuous phase transition to a BH state,\footnote{In the supersymmetric D3-D7 model, the careful analysis of \cite{Faulkner:2008hm} found that the analogous transition from MN to BH embeddings occurring at zero charge density was in fact third order.  For the D2-D8' system, we have numerically verified that the continuous phase transition on the increasing $d$ side of the MN state is also 3rd order.} but decreasing $d$ is discontinuous.  There is a narrow region with $d < \frac{3}{2} h$ where the charge susceptibility 
$\frac{\partial^2 F}{\partial d^2} < 0$ implying that the system is thermodynamically unstable.\footnote{Viewing the systems in the grand canonical ensemble, as the chemical potential is lowered, there is a first-order phase transition where the $d$ jumps discontinuously down from $\frac{3}{2} h$.}  

We will study further the nature of the transition from quantum Hall to metallic phases by examining the spectrum of fluctuations. In particular, will see this instability again in Sec. \ref{sec:hydromode} as a diffusion mode with a negative diffusion constant.

\subsection{Temperature-independent scaling}
\label{sec:scaling}
For the numerical analysis of the fluctuations, it is convenient to transform to a compact radial coordinate $\rho$ defined as
\be
\rho = \frac{u_T}{u} \ .
\ee
The boundary is at $\rho = 0$, and the horizon is now at $\rho = 1$.  In this coordinate, the location of the horizon, and thus the temperature, has been scaled out. 

We can completely scale out the temperature by defining scaled quantities as follows
\bea
\hat m &=& m \ u_T^{-1}\\
\hat d &=& \tilde d \ u_T^{-\frac{5}{2}} \\
\hat h &=& h \ u_T^{-\frac{5}{2}}\\
\hat b &=& b \ u_T^\frac{1}{2}\\
\hat a_0 &=& a_0 \ u_T^{-1}\\
\hat g &=&  g \ u_T^{-\frac{5}{2}}
\eea
and also rescaling the spacetime coordinates as $(\hat t, \hat x, \hat y) = u_T^{3/2} (t,x,y)$.  Using these scaled quantities, the action becomes
\be
S = - \mathcal N u_T^{7/2}\  \int d\rho  \left(\rho^{-9/2} \sqrt{A G H}  + c(\psi) \hat h \hat a_0' \right) \ ,
\ee
where now $' = \partial_\rho$ and
\bea
A(\rho) &=&1+f \rho^2\psi'^2-  \rho^4 \hat a_0'^2\\ 
G(\rho) &=&  \cos^6\psi \left(\hat b^2\rho  + \sin^4\psi\right) \\ 
H(\rho) &=& 1+\rho^5 \hat h^2 \quad .
\eea 
The scaled equations of motion are
\bea
 \hat a_0' &=& -\frac{f}{\rho^2 \hat g H} \hat d \\
 \partial_\rho \left(\hat g H \psi' \right) &=& \frac{f}{2\hat g \rho^7}\partial_\psi G + \hat h \hat a_0' \partial_\psi c 
\quad .
\eea

\section{Fluctuations}
\label{sec:fluctuations}

\subsection{Expansion}
We now allow for fluctuations of the following worldvolume fields:
\bea
\psi     	    & = & \bar\psi(\rho)+\epsilon \, \delta \psi(\hat t, \hat x,\hat y,\rho) \\
\hat a_0         & = & \bar a_0(\rho)+\epsilon \, \delta \hat a_0(\hat t, \hat x,\hat y,\rho) \\
\hat a_x         & = & \epsilon \, \delta \hat a_x (\hat t, \hat x,\hat y,\rho) \\
\hat a_y         & = & \hat h  \hat x +\epsilon \, \delta \hat a_y (\hat t, \hat x,\hat y,\rho) \\
\hat a_\rho        & = & \epsilon \, \delta \hat a_{\rho} (\hat t, \hat x,\hat y,\rho) \ ,
\eea
with $\epsilon$ a small parameter.  We do not consider fluctuations in the internal spheres or dependence on the internal $S^2$ or $S^3$ coordinates, thus restricting to the s-wave.

The Lagrangian can then be expanded to second order in $\epsilon$:
\be
 \mathcal L = \mathcal L_0 + \epsilon\mathcal L_1 + \epsilon^2\mathcal L_2 + \ldots \ .
\ee
Here $\mathcal L_0$ corresponds to the background Lagrangian, and $\mathcal L_1$ vanishes on-shell by the background equations of motion.  We derive the equations of motion for $\delta \hat\psi$, $\delta \hat a_0$, $\delta \hat a_x$,  and $\delta \hat a_y$ from $\mathcal L_2$.  In addition, since we are working in the radial gauge $\hat a_\rho = 0$, the equation of motion for $\delta \hat a_\rho$ becomes a constraint to be imposed on the fluctuations.

We make the following wavelike ansatz for the fluctuations:
\bea
  \delta \psi(\hat t,\hat x,\rho) & = & \delta \tilde\psi(\rho) e^{-i\hat \omega \hat t+i\hat k \hat x} \\
  \delta \hat a_\mu(\hat t,\hat x,\rho)  & = & \delta \tilde a_\mu(\rho) e^{-i\hat \omega \hat t+i\hat k \hat x}\ ,
\eea
where, because of the rotational symmetry in the $(x,y)$-plane, we can choose the momentum to be in the $x$-direction without loss of generality.

We must be careful to consider only physical fluctuations rather than gauge artifacts.  To ensure this, we work with the fluctuation of the longitudinal electric field:
\be
\delta \tilde e_x = \hat k \delta\tilde a_0 + \hat \omega \delta \tilde a_x 
\ee
which is gauge invariant.  As long as we start with initial conditions in radial gauge at some $\rho$ which satisfy the constraint given by the equation of motion for $\delta a_\rho$, the constraint and gauge condition will be satisfied by the solution at all $\rho$.

Note that we have introduced a rescaled frequency and momentum which are related to their unscaled counterparts as $(\hat\omega, \hat k) = u_T^{-3/2} (\omega, k)$.  As a result, the longitudinal electric field then scales as $\delta\tilde e_x = \delta e_x u_T^{-5/2}$.

We finally obtain a coupled system of linear, second order equations for the fluctuations.  The $\delta\tilde\psi$ equation of motion is
\bea 
\label{psiEOM}
&&\partial_\rho\left(\sqrt{\frac{GH}{A^3\rho^5}}f \left(\rho^4\bar\psi'\bar a_0'\delta\tilde a_0' + \left(1+\rho^4 \bar a_0'^2\right)\delta\tilde\psi' \right)\right)  
+\delta\tilde\psi \partial_\rho\left( \sqrt{\frac{GH\rho^5}{A}} \frac{\partial_{\bar \psi} \log G}{2}f\bar\psi' \right) 
\nonumber\\
&&- i\delta\tilde a_y \partial_\rho\left(\sqrt{\frac{G\rho^5}{AH}}f \hat h\hat k\bar\psi'\right) 
\nonumber\\
&&
=
 i\hat k\delta\tilde a_y \left(\partial_{\bar \psi}  c\bar a_0' - \sqrt{\frac{GA\rho}{H}} \frac{\partial_{\bar \psi}  \log G}{2}\hat h \right) 
-\delta\tilde\psi \Bigg( a_0'  \partial_{\bar \psi} ^2 c \nonumber\\ 
&&+ \sqrt{\frac{G\rho^3}{AH}} \left( \frac{H\hat\omega^2}{f} - k^2\left(1 - \rho^4 \hat k^2\bar a_0'^2 \right)\right)- \sqrt{\frac{AGH}{\rho^9}} \left(\frac{\partial_{\bar \psi}  \log G}{2}\right)^2 + \sqrt{\frac{AH}{G\rho^9}}\frac{\partial_{\bar \psi} ^2 G}{2}  \Bigg)
\nonumber\\
&& - i \sqrt{\frac{G\rho^5}{AH}}\hat k\bar\psi'\bar a_0'\delta\tilde e_x 
 - \delta\tilde a_y' \left(\hat h \partial_{\bar \psi}  c - \sqrt{\frac{GH}{A\rho}} \frac{\partial_{\bar \psi}  \log G}{2}\bar a_0' \right) \ .
\eea 
The equation of motion for $\delta\tilde a_0$ is
\bea
\label{a0EOM}
i\delta\tilde a_y \partial_\rho\left( c + \sqrt{\frac{G\rho^9}{AH}}\hat h \bar a_0'\right)
- \frac{\partial_\rho J}{\hat k}
 = \sqrt{\frac{G\rho^5}{AH}} \hat k\bar a_0' \bar\psi' \delta\tilde\psi + 
 \frac{i}{f}\sqrt{\frac{G\rho}{AH}}\left(1+f\rho^2\bar\psi'^2\right) \delta\tilde e_x \ ,
\eea
where we have defined the function $J$ as
\bea
\label{Jdefinition}
J(\rho) &=& \left(\hat h \partial_{\bar\psi} c + \sqrt{\frac{GH}{A\rho}} \frac{\partial_{\bar \psi}  \log G}{2}\bar a_0' \right) \delta\tilde\psi 
+ \sqrt{\frac{GH}{A^3\rho}}\left(1+f\rho^2\bar\psi'^2\right)\delta\tilde a_0'\nonumber\\
 &&- \sqrt{\frac{GH\rho^3}{A^3}} f \bar a_0' \bar\psi' \delta \tilde\psi' \ .
\eea
The equation for $\delta\tilde a_x$ is
\bea
\label{axEOM}
&&i\delta\tilde a_y \partial_\rho\left( c + \sqrt{\frac{G\rho^9}{AH}}\hat h \bar a_0'\right)
 +\partial_\rho\left( \sqrt{\frac{G}{AH\rho}} \frac{f \delta\tilde a_x'}{\hat\omega} \right) \nonumber\\
 &=& \sqrt{\frac{G\rho^5}{AH}} \hat k\bar a_0' \bar\psi' \delta\tilde\psi + 
 \frac{i}{f}\sqrt{\frac{G\rho}{AH}}\left(1+f\rho^2\bar\psi'^2\right) \delta\tilde e_x
\eea
and for $\delta\tilde a_y$ is
\bea
\label{ayEOM}
i\delta\tilde\psi \partial_\rho\left(\sqrt{\frac{G\rho^5}{AH}}f \hat h\hat k\bar\psi'\right)
-\delta\tilde e_x \partial_\rho\left(c + \sqrt{\frac{G\rho^9}{AH}}\hat h \bar a_0'\right)
+ \partial_\rho\left(\sqrt{\frac{G}{AH\rho}}f\delta\tilde a_y'\right) \nonumber\\
= i\hat k \left(-c\bar a_0' + \sqrt{\frac{GA\rho}{H}} \frac{\partial_{\bar \psi}  \log G}{2}\hat h \right) \delta\tilde\psi
+   \sqrt{\frac{G\rho}{AH}}\left( \frac{A}{H}\hat k^2 - \left(1+f\rho^2\bar\psi'^2\right) \frac{\hat\omega^2}{f} \right)\delta\tilde a_y
\eea
Finally, the equation derived by varying $\delta\tilde a_\rho$ is a constraint which imposes the gauge condition $a_\rho = 0$:
\bea
\label{arhoEOM}
J  +\sqrt{\frac{G}{AH\rho}} \frac{\hat k}{\hat\omega} \delta \tilde a_x'  =0\ .
\eea
The constraint can be used to eliminate one of the fluctuations in terms of the others.  Note that using the constraint, the $\delta\tilde a_0$ equation (\ref{a0EOM}) and the $\delta\tilde a_x$ equation (\ref{axEOM}) are identical, implying that there are only three independent equations of motion.  

\subsection{Numerical methodology}
Our goal is to solve the equations of motion for the fluctuations and find the normalizable modes.  The condition for normalizability is that the fluctuations vanish at the boundary:
\be
\delta \tilde e_x(0) = \delta \tilde a_y(0) = \delta \tilde\psi(0) = 0 \ .
\ee
At the horizon, the fluctuations have infalling boundary conditions.

In practice, to numerically solve the coupled equations for the fluctuations at fixed values of $\hat\omega$ and $\hat k$, we vary the initial conditions at the tip and shoot out to the boundary to look for solutions which are normalizable.  We must solve three coupled, linear, second-order differential equations for three independent functions.

An efficient technique for finding such solutions is the determinant method \cite{Amado:2009ts, Kaminski:2009dh}.  Rather than attempting to scan, for each $\hat\omega$ and $\hat k$, the full three-dimensional space of initial conditions, we choose a basis of linearly-independent initial conditions,
\be
\label{basis}
(\delta \tilde\rho(1), \delta \tilde e_x(1), \delta \tilde a_y(1)) = \left\{(1,0,0), (0,1,1), (0,1,-1)\right\} \ , 
\ee
for which we numerically solve the equations of motion.  We then compute the following determinant of boundary values:
\be
 \det\left( \begin{array}{ccc}
  \delta \tilde \psi^I & \delta\tilde \psi^{II} & \delta\tilde \psi^{III} \\
 \delta\tilde e_x^I & \delta\tilde e_x^{II} & \delta\tilde e_x^{III} \\
 \delta\tilde a_y^I & \delta\tilde a_y^{II} & \delta\tilde a_y^{III} 
\end{array} \right)\Bigg|_{\rho\to 0} \ ,
\ee
where the superscripts refer to the basis elements of (\ref{basis}).  If the determinant vanishes, there is at least one linear combination of the basis vectors for which $\delta\tilde\psi(0)$, $\delta\tilde e_x(0)$, 
and $\delta\tilde a_y(0)$ simultaneously vanish, signaling the existence of a normalizable mode.  
We repeat this procedure for various values of $\hat\omega$ and $\hat k$; for $\hat k=0$, a scan in $\hat\omega$ yields the spectrum, and allowing $\hat k > 0$ enables us to compute the dispersions.

For the figures presented in Sec.~\ref{sec:results}, we have chosen particular values of the parameters $u_T$, $m$, $d$, $h$, and $b$ for illustrative purposes.  As discussed in Sec.~\ref{sec:scaling}, the system is invariant under a scaling of the coordinates and parameters so there is, in fact, only a four-dimensional parameter space of BH solutions.  We used this rescaling symmetry to eliminate $u_T$ from the equations.  However, it makes more physical sense to work at fixed mass $m$ and background flux $b$, and vary the exogenous variables $u_T$, $h$, and $d$.  With the exception of Fig \ref{fig:kvsm}, which shows the dependence of the instability on the mass, all of the plots in Sec.~\ref{sec:results} are at fixed $m=1$ and $b=1$.

\section{Numerical results}
\label{sec:results}

\subsection{Hydrodynamical diffusion mode}
\label{sec:hydromode}

\begin{figure}[ht]
\center
\includegraphics[width=0.45\textwidth]{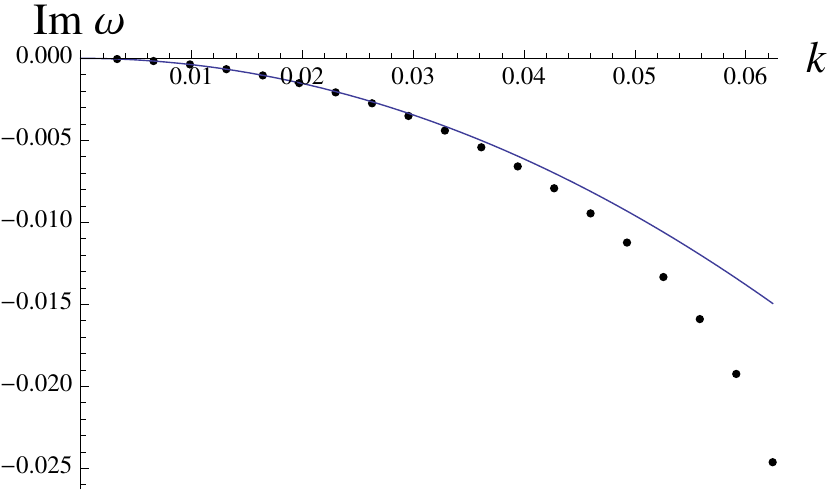}
\includegraphics[width=0.45\textwidth]{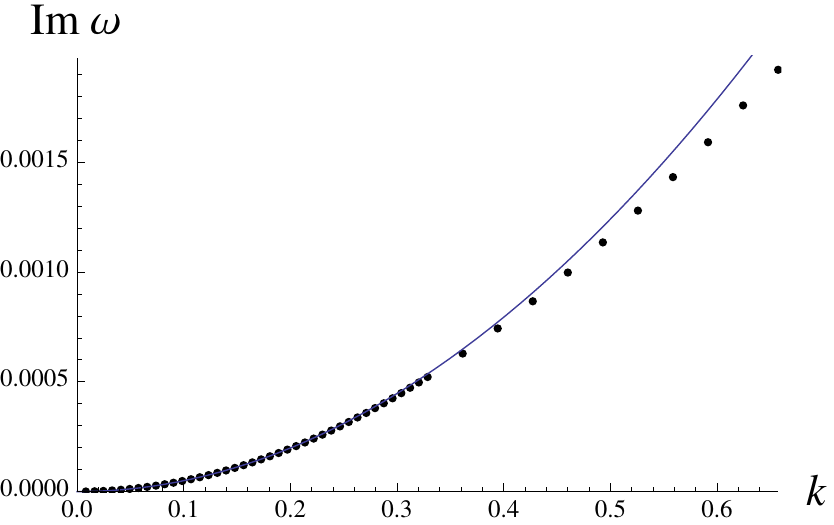}
\caption{The dispersion of the hydrodynamical mode at $h=0$ (left) and $h=0.35$ (right) for $u_T=0.3$, $d=0.2$, $m=1$, and $b=1$.  The curve is a fit to the quadratic dispersion (\ref{hydrodispersion}); from this fit, for $h=0$, $\Delta = 3.83$, and for $h=0.35$, $\Delta = - 4.96 \times 10^{-3}$.  A negative $\Delta$ indicates that the system is unstable to charge density fluctuations.}
\label{fig:hydrodispersion}
\end{figure}

The dominant quasinormal mode is the excitation with the smallest imaginary part, which then decays the slowest.  At long wavelengths, that is $k \to 0$, we find the longest lived mode is a purely imaginary, hydrodynamical diffusion mode, as shown in Fig.~\ref{fig:hydrodispersion}.  This mode has a dispersion 
\be
\label{hydrodispersion}
\omega = -i \Delta k^2 + \mathcal O(k^3) \ ,
\ee
where $\Delta$ is the diffusion constant.  This type of diffusion mode is generically found in systems with charge conservation; a local charge excess can only dissipate by diffusing through the medium.  Such a mode has been seen in other similar brane models, such as D3-D7 and D3-D7' \cite{Kaminski:2009dh, Bergman:2011rf, Davison:2011ek}. 

\begin{figure}[ht]
\center
\includegraphics[width=0.48\textwidth]{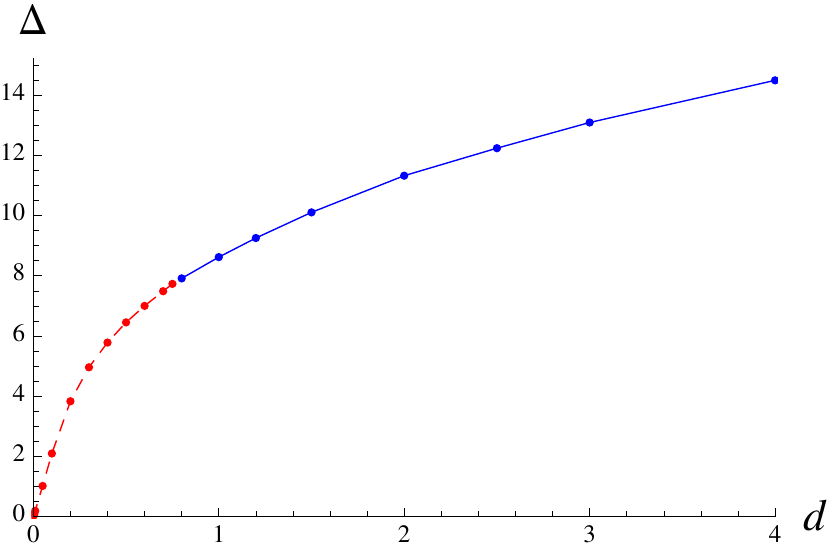}
\includegraphics[width=0.48\textwidth]{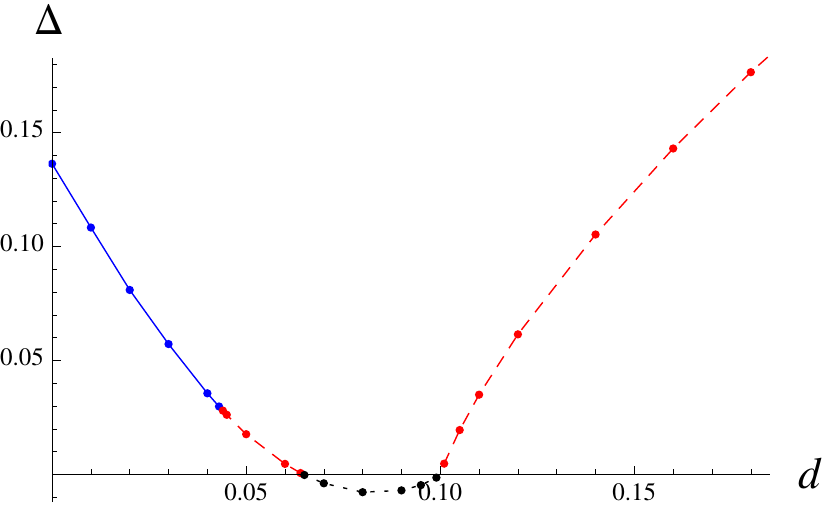}
\caption{The diffusion constant $\Delta$ of the hydrodynamical mode as a function of charge density $d$, (left) at zero magnetic field $h$ and (right) at magnetic field $h=0.15$, with  $u_T=0.3$, $m=1$, and $b=1$.  The solid blue portion of the curve indicates that the system is stable.  The black section where $\Delta <0$ is unstable.  The dashed red regions are subject to a modulated instability; see Sec.~\ref{sec:modulated_instabilities}.  }
\label{fig:diffusionvsd}
\end{figure}

\begin{figure}[ht]
\center
\includegraphics[width=0.48\textwidth]{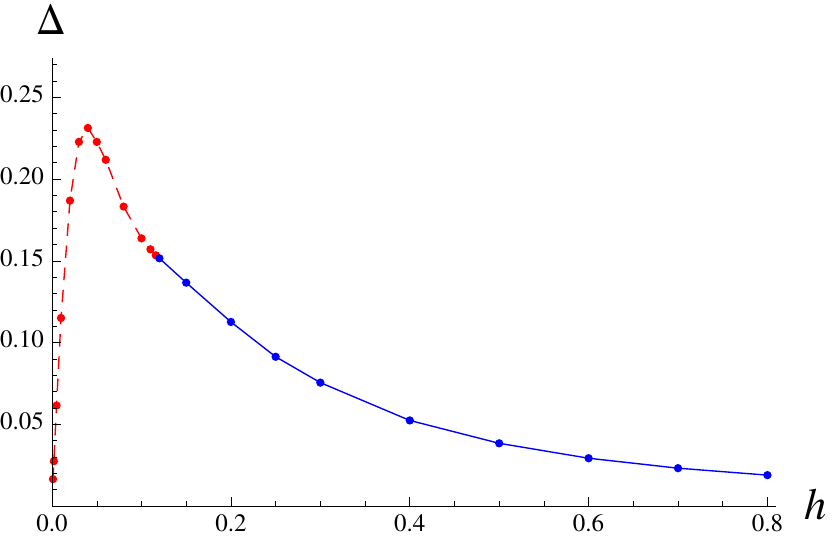}
\includegraphics[width=0.48\textwidth]{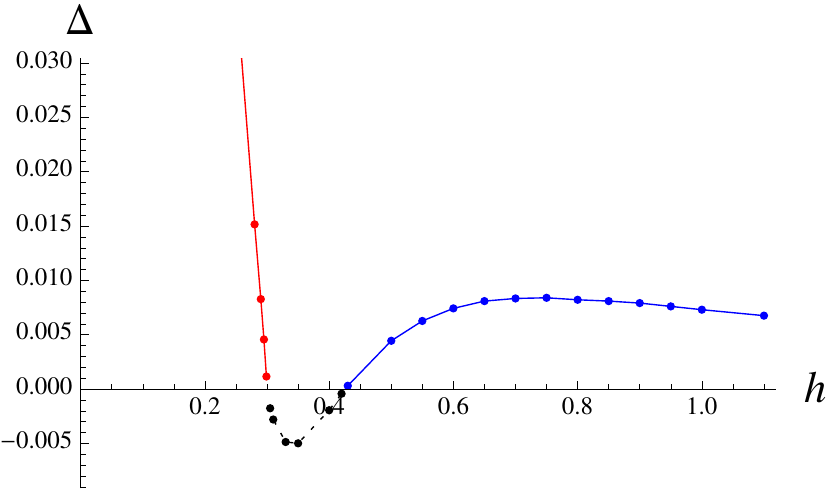}
\caption{The diffusion constant $\Delta$ plotted versus magnetic field $h$, (left) at $d=0$ and (right) at $d=0.2$, with  $u_T=0.3$, $m=1$, and $b=1$.  The solid blue part is stable, the black part is unstable, and the dashed red part is unstable, but only at nonzero momentum; see Sec.~\ref{sec:modulated_instabilities}.}
\label{fig:diffusionvsh}
\end{figure}

When the magnetic field is turned off, the diffusion constant grows with charge, as shown in Fig.~\ref{fig:diffusionvsd} (left).  In a nonzero magnetic field, Fig.~\ref{fig:diffusionvsd} (right), $\Delta$ similarly grows for $d> \frac{2}{3}h$.  At exactly $d= \frac{2}{3}h$, $\Delta = 0$; the system is in the gapped quantum Hall state, and the diffusion mode disappears.

At vanishing charge density, as the magnetic field is increased, $\Delta$ initially increases to a maximum and then decreases asymptotically to zero, as illustrated in Fig.~\ref{fig:diffusionvsh} (left).  At nonzero $d$, for some $h$ sufficiently larger than $\frac{3}{2} d$, $\Delta$ similarly reaches a maximum and then decays to zero; see Fig.~\ref{fig:diffusionvsh} (right), although the decay is not shown.  For $h$ below $\frac{3}{2} d$, $\Delta$ is decreasing.

\begin{figure}[ht]
\center
\includegraphics[width=0.48\textwidth]{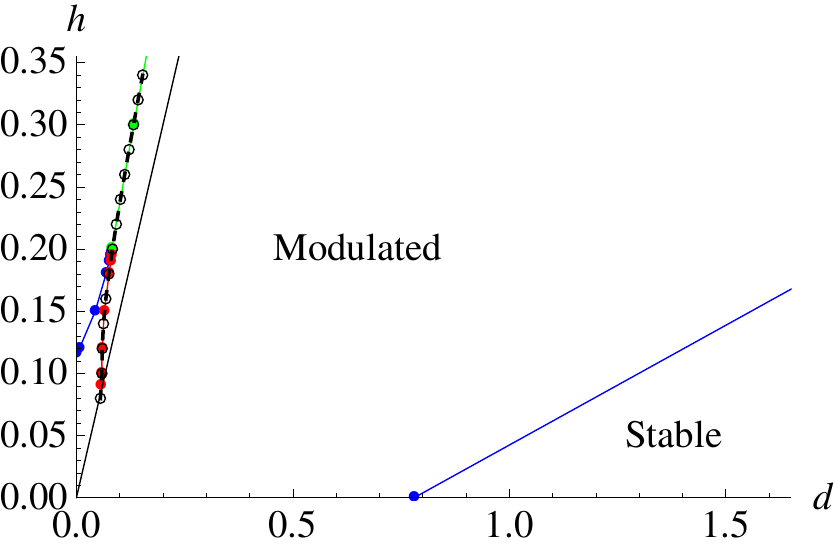}
\includegraphics[width=0.48\textwidth]{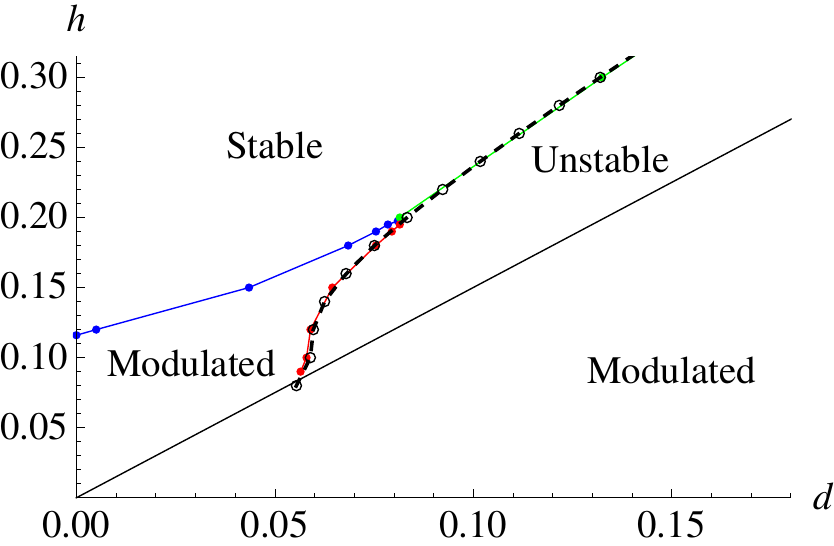} 
\caption{The unstable regions in the $(h,d)$-plane with $u_T=0.3$, $m=1$, and $b=1$.  
(Left) Overall plot showing all phases. (Right) A zoom into the region of small $d$ and $h$.  The straight black line, $h = \frac{3}{2} d$ is the stable QH state.  The region above that,
best seen in the right hand plot and labeled ``Unstable", is the region of thermodynamic instability.  The red curve is where $k_{min}$ hits zero.  On the green curve $\Delta$ crosses zero, and at the blue curves $k_{min} = k_{max}$. The dashed black curve with open boxes denotes the boundary of the region where $\frac{\partial \mu}{\partial d} < 0$ and coincides to a very high accuracy with the red and green curves.}
\label{fig:phase_diagram}
\end{figure}

However, the most striking features of both Figs.~\ref{fig:diffusionvsd} and \ref{fig:diffusionvsh} are the regions, drawn in black, where $\Delta < 0$, as illustrated in Fig.~\ref{fig:hydrodispersion} (right); this implies that $\rm{Im} \ \omega > 0$ and the diffusion mode is unstable.   Physically, this instability means that long wavelength fluctuations in the local density grow rather than diffuse away.  This is actually just another indication of the thermodynamic instability discussed in Sec.~\ref{sec:background}.  In fact, as can be seen in Fig.~\ref{fig:phase_diagram}, the narrow region adjacent to the quantum Hall state where $\Delta < 0$ matches precisely with the region where $\frac{\partial^2 F}{\partial d^2} = \frac{\partial \mu}{\partial d} < 0$.

\subsection{Zero sound}
\label{sec:zero-sound}
A generic consequence of a Fermi surface is the existence of a zero sound mode, which is a collective excitation of the Fermi surface.  Landau-Fermi liquid theory makes precise predictions for the properties of the zero sound.  At zero temperature,  the dispersion is
\be
\label{zerosounddispersion}
\omega \sim \pm c_s k - i \Gamma_q \ .
\ee
The zero sound speed $c_s = \frac{1}{\sqrt{d}}$, where $d$ is the number of spatial dimensions. The quantum attenuation  $\Gamma_q \sim k^2/\mu$ is due to decay into quasiparticle-quasihole pairs.  At nonzero temperature, thermal fluctuations further attenuate the zero mode.  For $k\mu \gg T^2$ the system is collisionless, and the zero sound with dispersion (\ref{zerosounddispersion}) persists, although $\Gamma_q$ is temperature-dependent.  When $k\mu \ll T^2$, the system is hydrodynamical, and the zero mode is completely damped and disappears. 

\begin{figure}[ht]
\center
\includegraphics[width=0.7\textwidth]{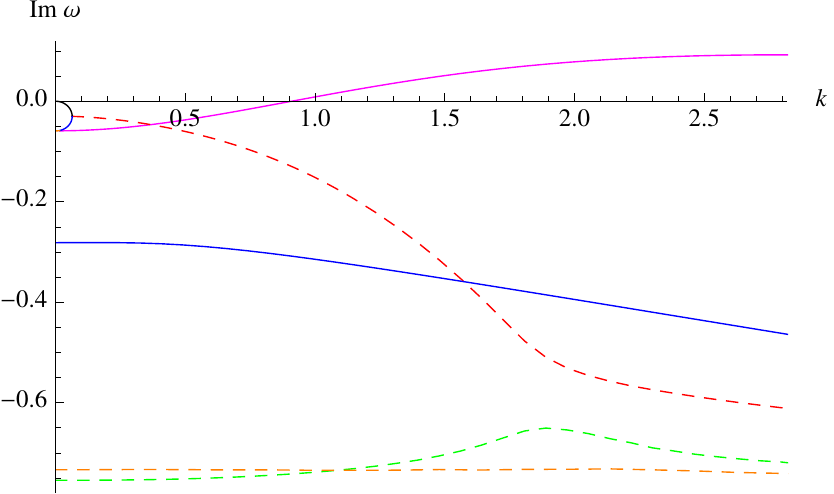}
\caption{The imaginary parts of the dispersions for the longest lived quasi-normal modes, for $d=0.2$, $h=0$, $u_T=0.3$, $m=1$, and $b=1$.  Purely imaginary modes are denoted by solid lines, while dashed curves represent complex modes.  The region near the origin is shown in more detail in Fig.~\ref{fig:dispersions_smallk} (top left).  The hydrodynamical diffusion mode, seen in Fig.~\ref{fig:hydrodispersion} (left), is the solid black curve.  The dashed red mode is the zero sound.  The solid magenta mode crosses the $k$-axis at $k = 0.91$, signaling the onset of a spatially modulated instability; see Sec.~\ref{sec:modulated_instabilities}.}
\label{fig:dispersions_imaginary}
\end{figure}

\begin{figure}[ht]
\center
\includegraphics[width=0.48\textwidth]{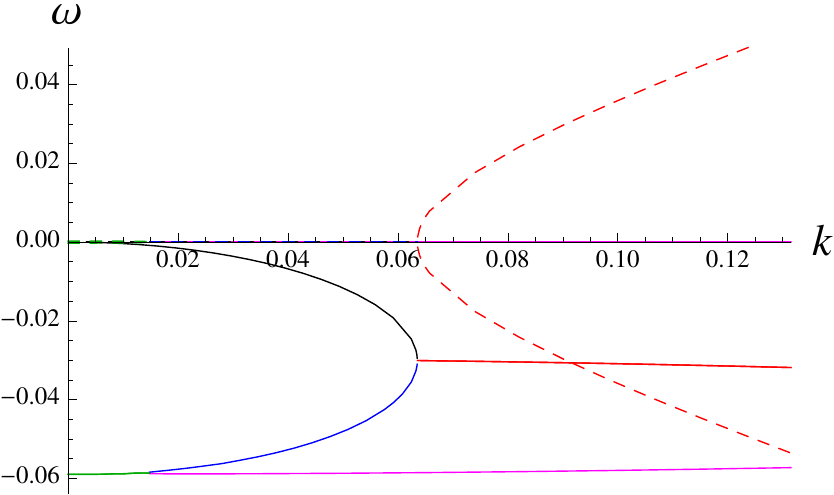}
\includegraphics[width=0.48\textwidth]{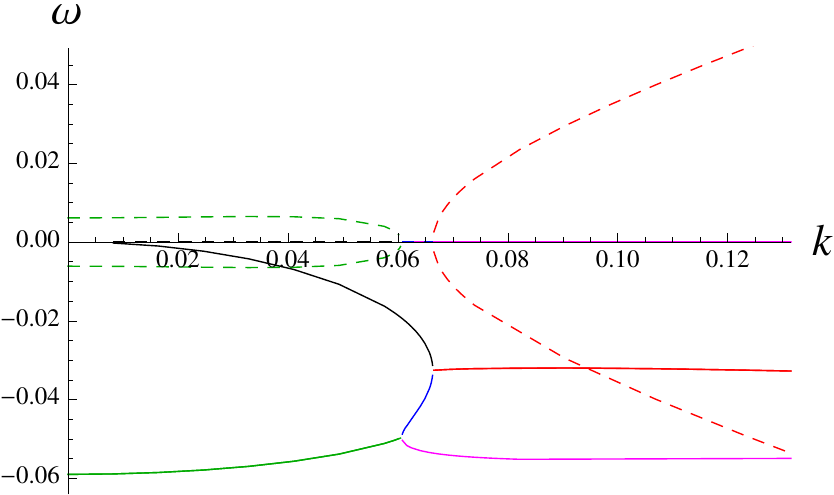}
\includegraphics[width=0.48\textwidth]{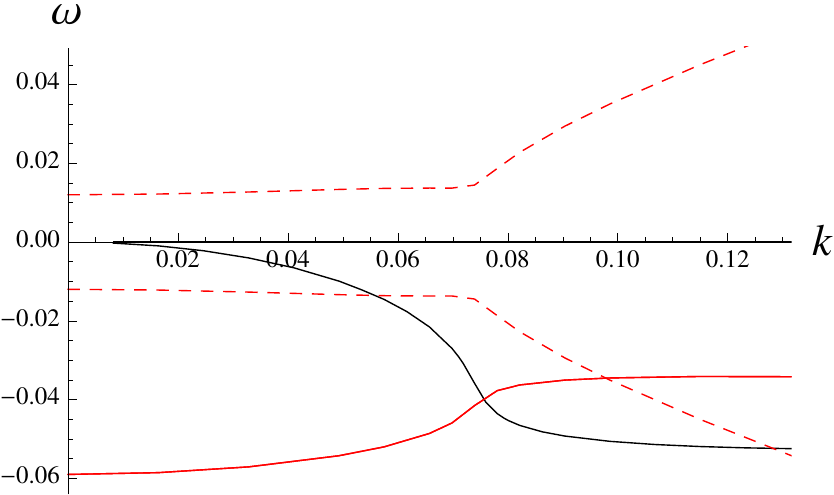}
\caption{The dispersions of the lowest lying quasi-normal modes at $h = 0$ (top left), 0.005 (top right), and 0.01 (bottom), with $d=0.2$, $u_T=0.3$, $m=1$, and $b=1$.  The real part of $\omega$ is shown by dashed curves, and the imaginary part by solid curves.  The purely dissipative hydrodynamic mode is black, and the zero sound is red.   (Top left and right) For $h < h_m \approx 0.01$, the hydrodynamic mode merges with a second, purely imaginary blue mode and then goes into the complex plane to become the zero sound.  The blue mode merges at lower $k$ with a magenta purely imaginary mode, becoming two green, long-wavelength, complex modes with nearly constant dispersion.  (Top right) This second merger point moves to larger $k$ as $h$ is increased.  (Bottom) Above the critical magnetic field $h > h_m$, the two merger points have met, leaving only the dissipative mode and the now massive zero sound.}
\label{fig:dispersions_smallk}
\end{figure}

Despite the lack of direct evidence for a Fermi surface, a holographic zero sound, which looks very much like the usual Fermi liquid zero sound, has been identified in a variety of D$p$-D$q$ systems \cite{Karch:2008fa, kp, Kim:2008bv, 1+1zerosound, Bergman:2011rf, Davison:2011ek, Jokela:2012vn, Goykhman:2012vy, Brattan:2012nb}.  The D2-D8' model is no exception, and the zero sound here shares many similarities with those of other holographic models.  In Fig.~\ref{fig:dispersions_imaginary}, we plot the imaginary part of the dispersion for the longest lived modes, at zero magnetic field.  Fig.~\ref{fig:dispersions_smallk} shows the dispersions of these low lying modes for various values of the magnetic field. 

At nonzero temperature and for small or vanishing magnetic field, the zero sound first appears at the transition from the hydrodynamical regime to the collisionless regime. At that point, the purely dissipative hydrodynamical mode merges with another purely imaginary mode; see Fig.~\ref{fig:dispersions_smallk} (top left).   The two modes then go off the imaginary axis into the complex plane to form the two branches of the zero sound with a nearly linear dispersion.  The location of the crossover from the hydrodynamical to collisionless regime is temperature-dependent: as the temperature is reduced, it occurs at smaller $k$, approaching $k=0$ as the temperature goes to zero.  This same low-frequency attenuation of the zero sound at nonzero temperature has been seen in other brane models \cite{Bergman:2011rf, Davison:2011ek, Brattan:2012nb}.

At small magnetic field, this second imaginary mode merges at a small $k$ with a third purely imaginary mode, forming another pair of complex modes.  For these long-wavelength modes, $\omega$ is nearly independent of $k$ and scales linearly with magnetic field.

\begin{figure}[ht]
\center
\includegraphics[width=0.7\textwidth]{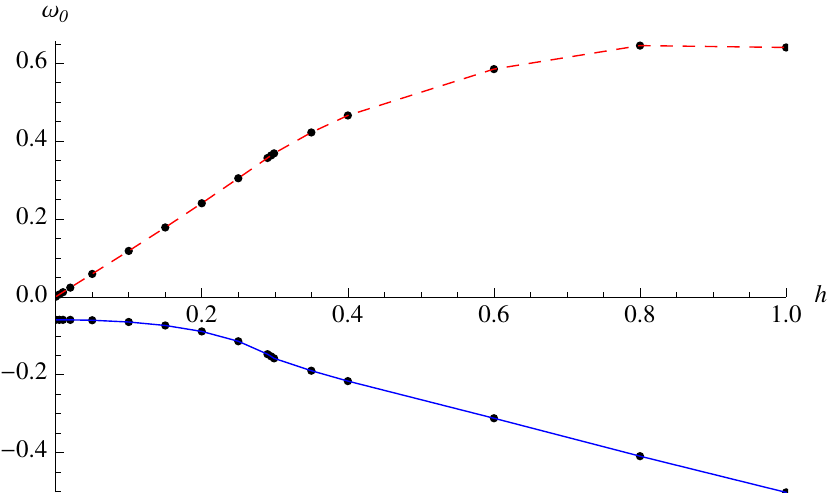}
\caption{The real (dashed red) and imaginary (solid blue) parts of $\omega$ for the zero sound mode at $k=0$ as a function of $h$, with $d=0.2$, $u_T=0.3$, $m=1$, and $b=1$.  The mass, given by the real part, increases linearly with magnetic field for $h\lesssim 0.3$.  
Notice that the region, where the merging of the modes depicted in Fig.~\ref{fig:dispersions_smallk} occurs, lies very close to the origin in this plot. }
\label{fig:zero-soundmass}
\end{figure}

As the magnetic field is increased, this second merger occurs at even larger $k$; 
see Fig.~\ref{fig:dispersions_smallk} (top right).  Eventually, at a critical magnetic field $h_m$, the two mergers meet, leaving a single, purely imaginary diffusion mode and a pair of complex 
modes, as shown in Fig.~\ref{fig:dispersions_smallk} (bottom).  We can identify the complex modes as the massive zero sound, whose mass grows linearly with $h$, at least for small $h$, as shown in Fig.~\ref{fig:zero-soundmass}.  Other probe brane models \cite{Jokela:2012vn, Goykhman:2012vy, Brattan:2012nb} also exhibit this type of magnetically-induced mass gap.

\subsection{Modulated instabilities}
\label{sec:modulated_instabilities}
\begin{figure}[ht]
\center
\includegraphics[width=0.7\textwidth]{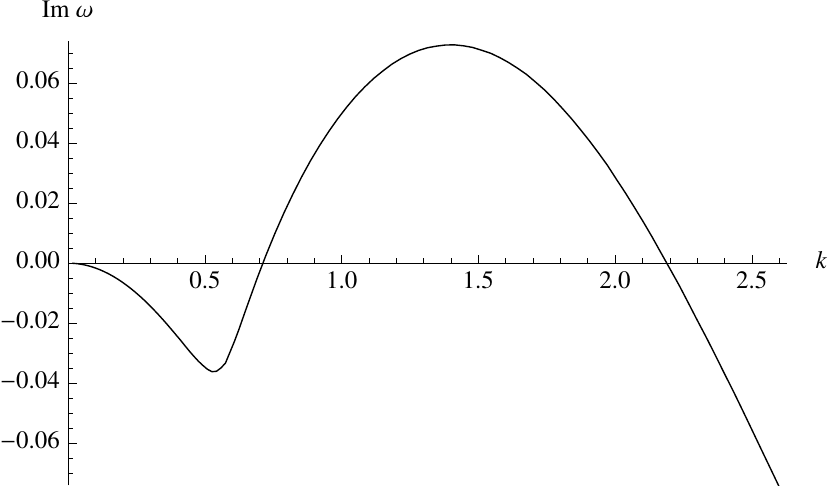}
\caption{The dispersion of the lowest purely imaginary mode with $d=0$, $h=0.1$, $u_T=0.3$, $m=1$, and $b=1$.  Between $k_{min} = 0.712$ and $k_{max} = 2.189$,  ${\rm Im} \, \omega > 0$ signaling the existence of a modulated instability.}
\label{fig:dispersion_instability}
\end{figure}

Following the lowest purely imaginary mode to larger $k$,  in some cases, ${\rm Im} \, \omega$ becomes positive for a range of momenta, signaling the onset of an instability; a typical such dispersion is shown in Fig.~\ref{fig:dispersion_instability}.  Because the unstable modes are at $k > 0$, this decay leads to spatial modulation, breaking translation invariance.  

This modulated instability is a variant of the type first investigated in \cite{Nakamura:2009tf} and subsequently seen in many other holographic contexts \cite{inhomo, Bergman:2011rf, Jokela:2012vn}.  Chern-Simons gauge theories are prone to modulated instabilities in the presence of electric fields but are typically stabilized by magnetic fields.  In many holographic contexts, the radial electric field dual to the charge density is responsible for the instability.  Here, the situation is somewhat more complicated, as the fluctuations are completely coupled at $k>0$, and the instability can persist down to zero density.

Roughly speaking, we find the instability occurs at low temperatures, large mass, and in the vicinity of the quantum Hall state.  That is, for fixed $h$ and $d$, there is a critical temperature $u_T^c$ above which the instability disappears, and sufficiently far from $d=\frac{3}{2} h$, the system is stable.  In addition, the unstable region of parameter space shrinks as the mass is reduced.  

\begin{figure}[ht]
\center
\includegraphics[width=0.48\textwidth]{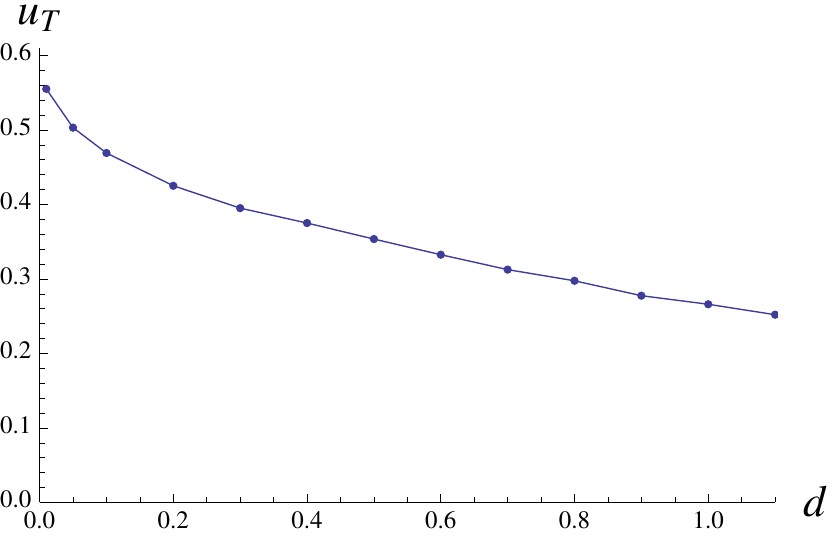}
\includegraphics[width=0.48\textwidth]{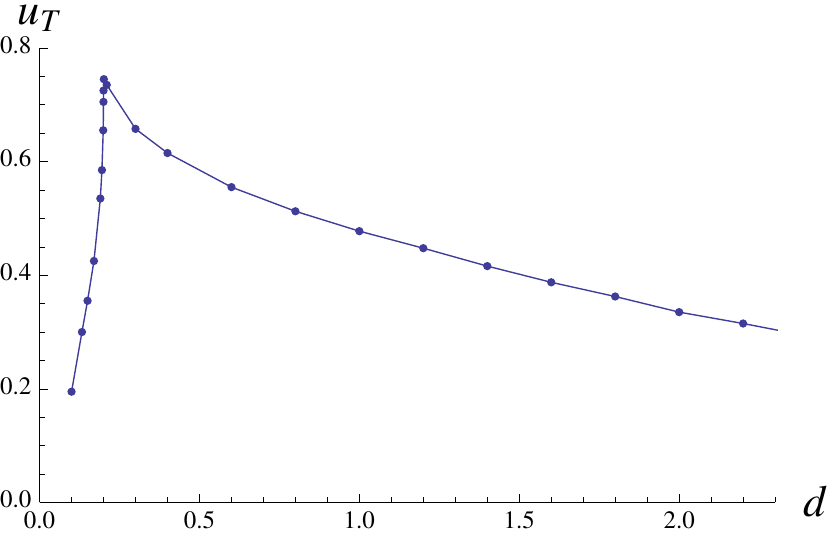}
\caption{The critical temperature $u_T$ below which the system suffers a modulated instability as a function of $d$, (left) for $h=0$ and (right) for $h=0.3$, with $m=1$, and $b=1$. }
\label{fig:uTvsd}
\end{figure}

\begin{figure}[ht]
\center
\includegraphics[width=0.48\textwidth]{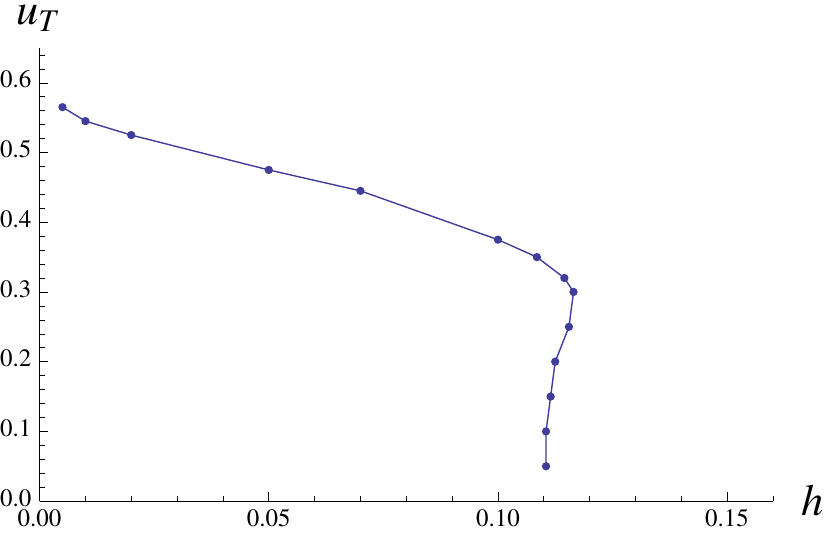}
\includegraphics[width=0.48\textwidth]{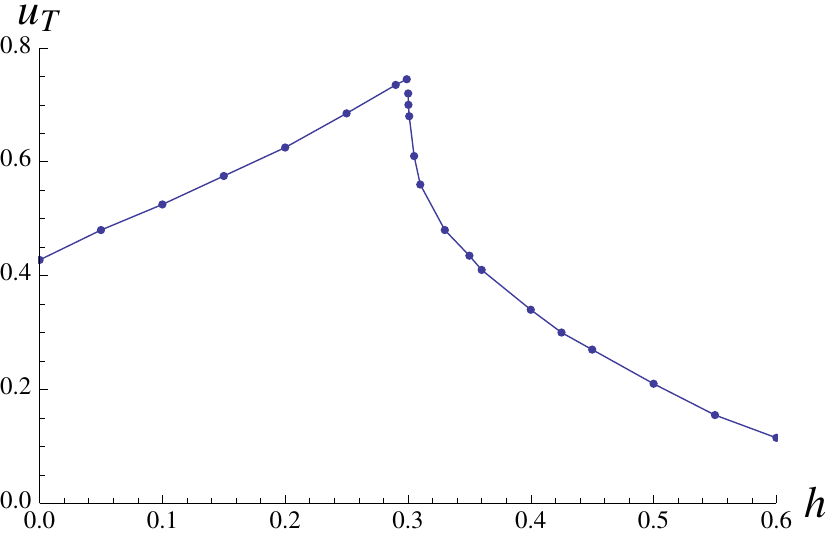}
\caption{The critical temperature $u_T$ versus magnetic field $h$ for (left) $d=0$ and (right) $d=0.2$, with $m=1$, and $b=1$. }
\label{fig:uTvsh}
\end{figure}

We can illustrate this for representative parameters in various ways.  Fig.~\ref{fig:phase_diagram} shows the region of the $(h,d)$-plane, at fixed temperature, where the modulated instability occurs. 
Fig.~\ref{fig:uTvsd} shows $u_T^c$ as a function of $d$ for two different values of $h$, while Fig.~\ref{fig:uTvsh} plots $u_T^c$ as a function of $h$ for two different values of $d$. 

The behavior of the critical temperature is not monotonic either as a function of density or magnetic field, but, surprisingly, it depends on which side of the quantum Hall state one is on.  For $h > \frac{2}{3} d$, $u_T^c$ is decreasing as $h$ increases, and, as expected, the magnetic field is stabilizing the system.  However, for $h < \frac{2}{3} d$, increasing $h$ raises $u_T^c$, evidently destabilizing the system.  Note, in particular, that for $d > \frac{3}{2} h$, increasing the density decreases the critical temperature.  This is in contrast to what was seen in the D3-D7' model \cite{Bergman:2011rf}, where $u_T^c$ goes down with increasing $d$.  
 
\begin{figure}[ht]
\center
\includegraphics[width=0.48\textwidth]{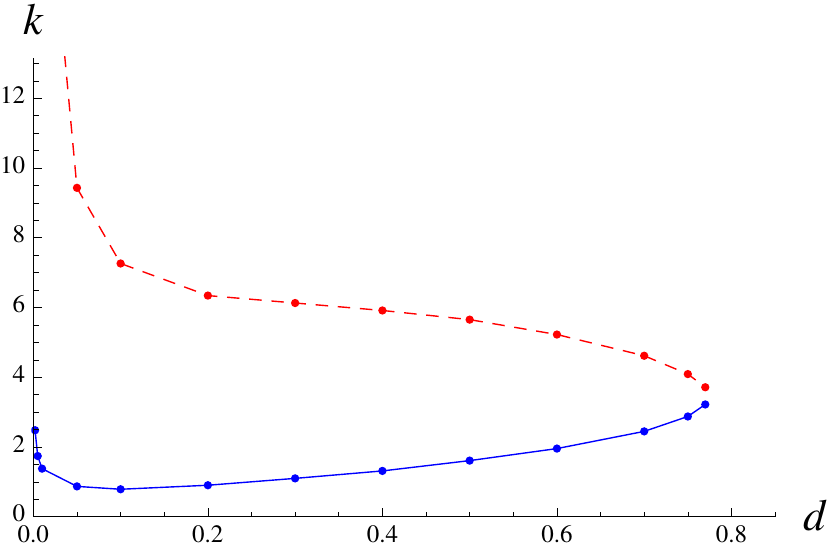}
\includegraphics[width=0.48\textwidth]{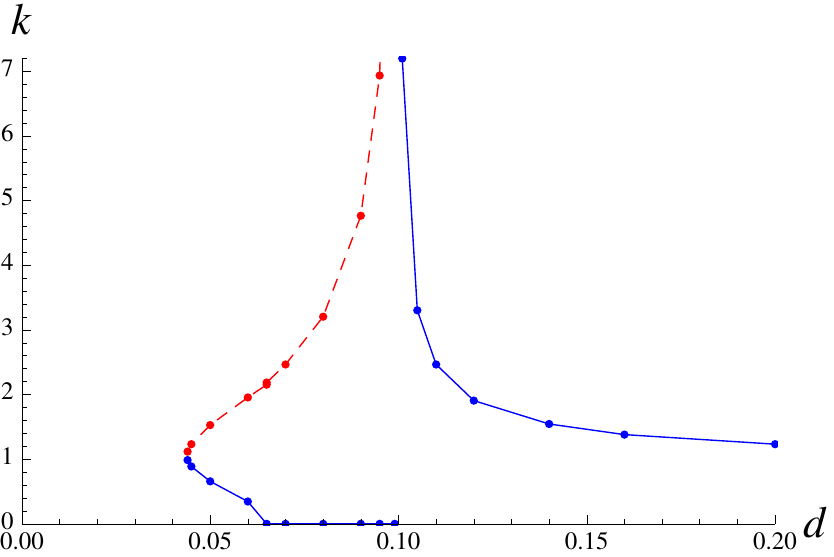}
\caption{The minimum (solid blue) and maximum (dashed red) momentum $k$ where the lowest purely imaginary mode is unstable as a function of charge density $d$, (left) at $h=0$ and (right) at $h=0.15$, with  $u_T=0.3$, $m=1$, and $b=1$.  Note that for $h=0.15$ (right), $k_{min}=0$ for $d \in (0.065,0.1)$, implying that there is a homogeneous instability. 
In the right hand plot, because we have zoomed in to show the behavior around $d \sim 0.1$, the maximum momentum is above the region shown; for $d \in (0.1, 0.2)$, $k_{max} \gtrsim 25$. }
\label{fig:kvsh}
\end{figure}

\begin{figure}[ht]
\center
\includegraphics[width=0.48\textwidth]{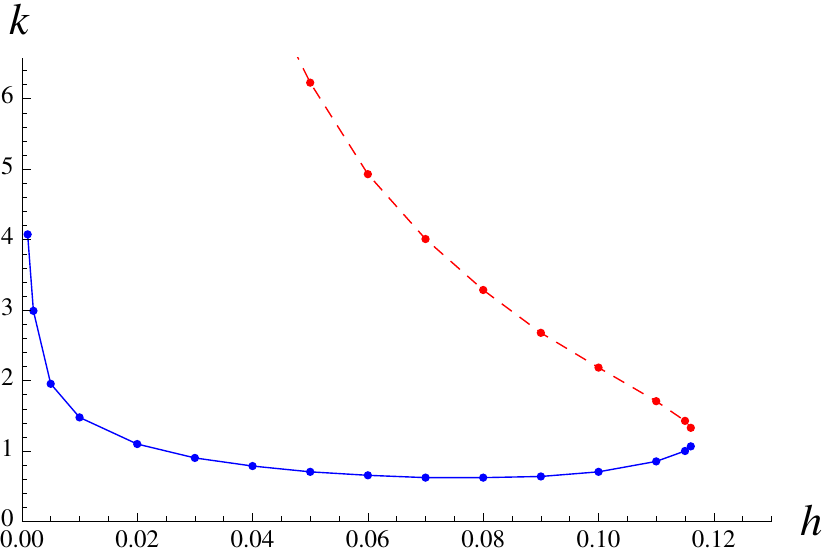}
\includegraphics[width=0.48\textwidth]{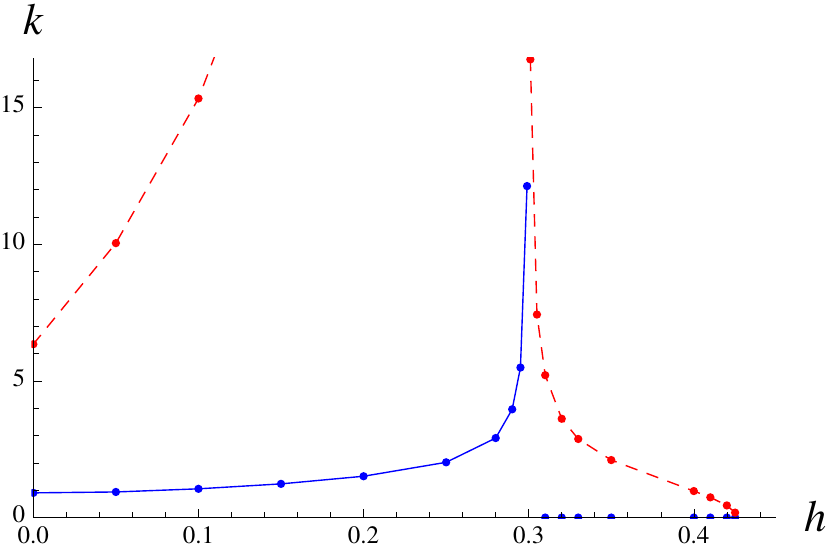}
\caption{The minimum (solid blue) and maximum (dashed red) momentum $k$ where the lowest purely imaginary mode is unstable as a function of charge density $h$, (left) at $d=0$ and (right) at $d=0.2$, with  $u_T=0.3$, $m=1$, and $b=1$.   For $d=0.2$ (right), $k_{min}=0$ and there is a homogeneous instability for $h \in (0.3,0.42)$.}
\label{fig:kvsd}
\end{figure}

We present in Figs.~\ref{fig:kvsh} and \ref{fig:kvsd} the range in $k$ of unstable modes for four slices across the $(h,d)$-plane shown in Fig.~\ref{fig:phase_diagram}.  Stability away from $d=\frac{3}{2} h$ occurs when $k_{min}=k_{max}$. The $k_{min}$ ($k_{max}$) is the minimum (maximum) momentum above (below) of which there is an unstable mode.  Both Figs.~\ref{fig:kvsh} and \ref{fig:kvsd} include domains where $k_{min}=0$; these correspond to regions of homogeneous thermodynamic instability.

Although there are unstable regions on either side of $d=\frac{3}{2} h$, the quantum Hall state itself is stable \cite{Jokela:2011sw}.  Counterintuitively, as can be seen from Figs.~\ref{fig:uTvsd} and Fig.~\ref{fig:uTvsh}, the critical temperature increases as the quantum Hall state is approached, implying that the system is becoming more unstable, rather than less.  However, as is evident from Figs.~\ref{fig:kvsh} and \ref{fig:kvsd}, both $k_{min}$ and $k_{max}$ grow toward infinity as $d=\frac{3}{2} h$ is approached.  The exception is when the quantum Hall state is approached from the thermodynamically unstable region, in which case $k_{min}$ stays zero.

\begin{figure}[ht]
\center
\includegraphics[width=0.45\textwidth]{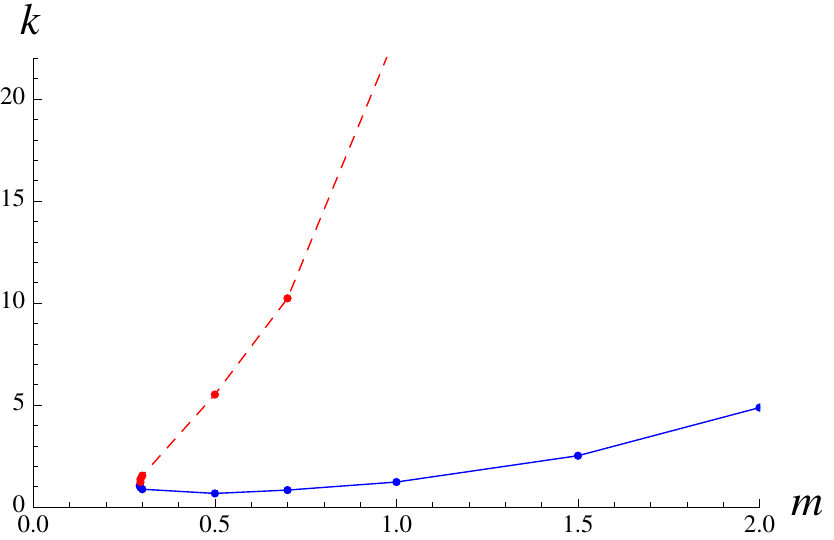}
\includegraphics[width=0.45\textwidth]{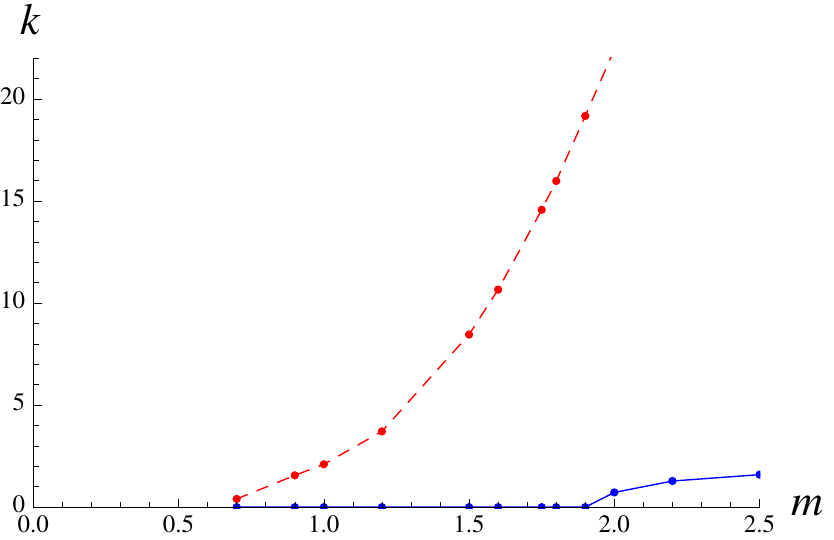}
\caption{The range in momenta, from $k_{min}$ (solid blue) and to $k_{max}$ (dashed red), for which the there is an instability plotted as a function of the mass $m$, (left) at $h=0.15$ and (right) at $h=0.35$, with  $u_T=0.3$, $d=0.2$, and $b=1$.   For $h=0.35$ (right), $k_{min}=0$ in the range $m \in (0.69,1.9)$  and there is a homogeneous instability.}
\label{fig:kvsm}
\end{figure}

Finally, we illustrate the effect of the mass on stability in Fig.~\ref{fig:kvsm}; we again plot the range in $k$ of unstable modes, this time as a function of $m$.  The range of unstable momenta shrinks as the mass is reduced.  For a given $d$, $h$, and $u_T$, there is a sufficiently small $m$ such that the modulated instability disappears entirely.

\section{Discussion}
\label{sec:discussion}

Our investigation of the fluctuations of the D2-D8' model found a quasi-normal mode spectrum in many ways qualitatively in line with those of other related probe brane systems.  We found a long-wavelength diffusion mode in the hydrodynamical regime and a holographic zero sound in the collisionless regime.  An applied magnetic field modifies the spectrum and, in particular, generates a mass for the zero sound.  At low temperature and for the charge density and magnetic field in a range near the quantum Hall state, the homogeneous phase suffers a spatially modulated instability.  In certain parameter regions the minimal spatial modulation goes to zero continuously, reproducing the thermodynamic instability found in \cite{Jokela:2011eb}.

We did, however, find some significant differences, in particular, with D3-D7' model to which the D2-D8' model is most closely related.  Most notably, we found here an unstable region near the quantum Hall state, while in the D3-D7' model, the instability was found only at high density and low magnetic fields.  
In addition, we found smaller masses led to improved stability, which is exactly the opposite of what was seen in the D3-D7' case.  Considering that the instabilities stem from essentially the same mechanism, this is quite surprising.  One potentially related issue is that in \cite{Bergman:2011rf, Jokela:2012vn}, the two independent fluxes of the D3-D7' model were chosen to be equal, which excluded the possibility of a  quantum Hall state.  If one of the fluxes were shut off, the unstable region of the D3-D7' could possibly also be centered around the now allowed quantum Hall state.

An important open problem is to find the ground state to which this modulated instability ultimately leads. As the unstable modes have nonzero $k$, it is reasonable to expect that translation invariance will be broken in the ground state.   Modulated states, such as spin and charge density waves, with spontaneously broken translation symmetry, appear in various condensed matter contexts.\footnote{For reviews of spin and charge density waves in condensed matter systems, see \cite{Gruner}.}   Explicit modulated ground states have been studied in some simpler holographic systems exhibiting modulated instabilities \cite{inhomo}.  In this model, finding such a state is technically more challenging and is a subject of ongoing investigation.  Because all the fields are coupled, we anticipate a striped phase in which all quantities, such as charge density, currents, and the fermion condensate, are spatially modulated.

As the unstable region was found in the vicinity of the QH state in this model, it would be interesting to check if the quasi-normal modes can be matched to the normal modes of the QH state in this region. In particular, one could expect that the zero sound approaches the lowest QH mode, which is a magneto-roton in a vast region of the parameter phase \cite{Jokela:2011sw, Jokela:2010nu}, as the magnetic field approaches the locking value of the QH phase. As the numerical analysis of the quasi-normal modes close to this phase is very involved, our attempts to clarify this issue remain inconclusive.

Another outstanding issue is the existence of a Fermi surface.  Fermi surfaces have been argued to exist in holographic quantum liquids, but no compelling evidence for them in probe brane models has been found.  It may be that if the charged fermions are holographically encoded in bulk degrees of freedom at the black hole horizon, the Fermi surface is fractionalized and difficult to detect by the usual gauge invariant probes.  For this reason, it may be useful to focus on those models with charged fermions dual to sources outside horizon.  In brane models with \#ND = 6, background magnetic fields can induce, via Chern-Simons terms, charge along the holographic radial direction. This effect has been studied in the D2-D8' \cite{Jokela:2011eb}, D3-D7' \cite{Bergman:2010gm}, and Sakai-Sugimoto models \cite{Lifschytz:2009si}, where the charge carried on the horizon and the charge spread radially along the brane behave as two fluids with distinct properties.  Perhaps, these can serve roughly as probe brane 
analogs of the electron star \cite{Hartnoll:2010gu,Puletti:2010de}.

We end this paper by discussing another possible outgrowth of this work. In the D3-D7' model, in \cite{Semenoff}, several non-trivial examples has been uncovered where the weak-coupling
results for certain observables has been matched with the results of holographic strong-coupling calculations. It would be interesting to study whether this analysis could be carried over also for the D2-D8' model. However, this would be much more involved in the present context due to background being only conformal to $AdS$ and the lack of massless probe embeddings.

\bigskip
\noindent

{\bf \large Acknowledgments}

We would like to thank Paolo Benincasa, Oren Bergman, Danny Brattan, Richard Davison, Simon Gentle, Gilad Lifschytz, and Andrei Parnachev for helpful discussions, comments, and suggestions.  
N.J. is supported by the MICINN and FEDER (grant FPA2008-01838), the
Spanish Consolider-Ingenio 2010 Programme CPAN (CSD2007-00042), and the Xunta de
Galicia (Conselleria de Educacion and grant INCITE09-206-121-PR). N.J. is
also supported by the Juan de la Cierva program.  
The work of M.J. is supported in part by EU grants PIEF-GA-2011-300984, PERG07-GA-2010-268246 and the EU program ``Thales'' ESF/NSRF 2007-2013.
It has also been co-financed by the European Union (European Social Fund, ESF) and
Greek national funds through the Operational Program ``Education and Lifelong Learning'' of the National Strategic
Reference Framework (NSRF) under ``Funding of proposals that have received a positive evaluation in the 3rd and 4th Call of ERC Grant Schemes''. 
The research of M.L. was supported by the European Union grant FP7-REGPOT-2008-1-CreteHEPCosmo-228644 and by the EU program ``Thales'' ESF/NSRF 2007-2013.  In addition, M.L. is also supported by funding from the European Research
Council under the European Union's Seventh Framework Programme (FP7/2007-2013) /
ERC Grant agreement n¡ 268088-EMERGRAV.
M.J. and M.L. would also like to thank the University of Santiago de Compostela, and N.J. and M.L. the Technion for warm hospitalities while this project was underway.

\appendix

\end{document}